\documentclass[11pt]{article}

\usepackage{amsmath} 
\usepackage{amssymb} 
\usepackage{graphicx} 

\usepackage[labelfont=bf,labelsep=period,justification=raggedright,font=small]{caption}

\usepackage{xspace}
\usepackage{color}

\DeclareMathOperator{\MyRe}{Re}
\DeclareMathOperator{\MyIm}{Im}
\DeclareMathOperator{\Tr}{Tr}

\date{\today}

\title{Low-dimensional firing rate dynamics of spiking neuron networks}

\author{Maurizio Mattia\\
\\
\footnotesize Istituto Superiore di Sanit\`a\\
\footnotesize viale Regina Elena 299, 00161 Rome, Italy\\
\footnotesize\tt maurizio.mattia@iss.it}

\begin{document}

\maketitle

\begin{abstract}
Starting from a spectral expansion of the Fokker-Plank equation for the membrane potential density in a network of spiking neurons, a low-dimensional dynamics of the collective firing rate is derived. As a result a $n$-order ordinary differential equation for the network activity can be worked out by taking into account the slowest $n$ modes of the expansion. The resulting low-dimensional dynamics naturally takes into account the strength of the synaptic couplings under the hypothesis of a not too fast changing membrane potential density. By considering only the two slowest modes, the firing rate dynamics is equivalent to the one of a damped oscillator in which the angular frequency and the relaxation time are state-dependent. The presented results apply to a wide class of networks of one-compartment neuron models.
\end{abstract}

The collective dynamics of neuronal networks can be complex even when simplified one-compartment models of neurons are considered for modeling. Complexity arises from the spiking sparse-in-time nature of the inter-neuronal communication, the high-dimensionality of the system and the quenched randomness in the synaptic couplings. Reducing such complexity relying on mean-field approaches has a long history in theoretical neuroscience \cite{Wilson1972,Amit1991,Gerstner1995,Shriki2003,Schaffer2013,Montbrio2015}, although adopted approximations often limit the general applicability of the resulting simplified dynamics. Here, with the aim to widen such effectiveness by relying on a population density approach \cite{Treves1993,Abbott1993,Knight1996,Brunel1999,Knight2000,Mattia2002}, a low-dimensional dynamic mean-field description is provided for the instantaneous emission/firing rate $\nu(t)$ of a network of spiking neurons. In this framework, the network dynamics is described by the Fokker-Planck (FP) equation for the membrane potential density. A suited spectral expansion of the FP operator \cite{Knight2000,Mattia2002} is the approach followed focusing on a homogeneous pool of interacting integrate-and-fire (IF) neurons.

\section*{Notation summary}

To start, some of the results in \cite{Mattia2002} and a brief description of the adopted notation are provided. 

Under mean-field approximation the density $p(v,t)$ of neurons  with membrane potential $v$ at time $t$ follows a FP equation with suited boundary conditions:
\begin{equation}
   \partial_t \, p(v,t) = L \, p(v,t) \equiv  - \partial_v \, S_p(v,t) \, .
\label{eq:FP}
\end{equation}
In general, the FP operator $L \equiv L(p)$ is nonlinear because it depends on the instantaneous firing rate $\nu(t)$ given by the flux of realizations crossing the emission threshold $\theta$:
\begin{displaymath}
   \nu(t) = S_p(\theta,t) \, .
\end{displaymath}

The spectrum $\{\lambda_n\}$ of the FP operator,
\begin{displaymath}
   L \, | \phi_n \rangle = \lambda_n \, | \phi_n \rangle \, ,
\end{displaymath}
provides a moving basis $\{| \phi_n \rangle\}$ driven by the first two instantaneous moments of incoming currents, which are time-varying and state-dependent. Equation~(\ref{eq:FP}) can be rewritten as the dynamics of the expansion coefficients $a_n(t)$ of the density $p(v,t)$ on such basis:
\begin{displaymath}
   | p \rangle = \sum_n a_n | \phi_n \rangle \, .
\end{displaymath}

An emission rate equation results:
\begin{equation}
\left\{
\begin{array}{rcl}
   \dot{\vec{a}} & = & ({\mathbf \Lambda} + {\mathbf C} \, \dot{\nu}) \, \vec{a} + \vec{c} \, \dot{\nu} \\
   \nu & = & \Phi + \vec{f} \cdot \vec{a}
\end{array}
\right.,
\label{eq:ERE}
\end{equation}
where $\vec{a} = \{ a_n \}_{n \neq 0}$. $\vec{f} =\{ S_{\phi_n}(\theta,t) \}_{n \neq 0}$ are the fluxes in $\theta$ for non-stationary modes ($n \neq 0$). Synaptic coupling in stationary ($n=0$) and non-stationary modes is expressed in the vector $\vec{c}$,
\begin{displaymath}
   c_n = \langle \partial_\nu \, \psi_n | \phi_0 \rangle
         \;\;\; \forall n \neq 0,
\end{displaymath}
and the matrix ${\mathbf C}$,
\begin{displaymath}
   C_{nm} = \langle \partial_\nu \, \psi_n | \phi_m \rangle
         \;\;\; \forall n,m \neq 0 \, ,
\end{displaymath}
where $\langle \psi_n |$ are the eigenfunctions of the adjoint operator $L^+$. ${\mathbf \Lambda}$ is the diagonal matrix of the eigenvalues of $L$
\begin{displaymath}
   \Lambda_{nm} = \lambda_n \, \delta_{nm} \;\;\; \forall n,m \neq 0.
\end{displaymath}

\section*{Low-dimensional ordinary differential equation for $\nu$}

\subsection*{A set of uncoupled IF neurons}

In absence of synaptic coupling, incoming currents to the neurons do not depend on $\nu(t)$. Hence, both $\vec{c} = 0$ and ${\mathbf C} = 0$, and eigenfunctions of $L$ are independent from $\nu(t)$. Under these conditions Eq.~(\ref{eq:ERE}) simplifies as
\begin{equation}
\left\{
\begin{array}{rcl}
   \dot{\vec{a}} & = & {\mathbf \Lambda}  \, \vec{a} \\
   \nu & = & \Phi + \vec{f} \cdot \vec{a}
\end{array}
\right. ,
\label{eq:EREuncoupled}
\end{equation}
with constant coefficients $\Phi$, $\vec{f}$ and ${\mathbf \Lambda}$.

Recursively deriving  with respect to time both equations in (\ref{eq:EREuncoupled}), one obtains:
\begin{displaymath}
\begin{array}{rcl}
   \dot{\nu} & = & \vec{f} \cdot \dot{\vec{a}} \\
   \ddot{\nu} & = & \vec{f} \cdot \ddot{\vec{a}} \\
            & \cdots  & \\
   \partial_t^n{\nu} & = & \vec{f} \cdot \partial_t^n{\vec{a}}
\end{array}
\end{displaymath}
and
\begin{displaymath}
   \partial_t^n \vec{a}  = {\mathbf \Lambda}  \, \partial_t^{n-1}{\vec{a}} = {\mathbf \Lambda}^2  \, \partial_t^{n-2}{\vec{a}} = \cdots = {\mathbf \Lambda}^n  \, \vec{a} \, ,
\end{displaymath}
such that
\begin{displaymath}
   \partial_t^n \nu  =  \vec{f} \cdot {\mathbf \Lambda}^n  \, \vec{a} \, .
\end{displaymath}

Stopping derivations to the $n$-th order, an approximated expression for the emission rate equation results:
\begin{displaymath}
   \nu = \Phi + f_1 \, a_1 + f_2 \, a_2 + \cdots  + f_n \, a_n 
\end{displaymath}
and
\begin{displaymath}
\begin{pmatrix}
   \dot{\nu} \\
   \ddot{\nu} \\
   \vdots \\
   \partial_t^n \nu
\end{pmatrix}
=
\begin{pmatrix}
   \lambda_1 & \lambda_2 & \cdots & \lambda_n \\
   \lambda_1^2 & \lambda_2^2 & \cdots & \lambda_n^2 \\
   \vdots  & \vdots & \ddots & \vdots \\
   \lambda_1^n & \lambda_2^n & \cdots & \lambda_n^n
\end{pmatrix}
\begin{pmatrix}
   f_1 \, a_1 \\
   f_2 \, a_2 \\
   \vdots \\
   f_n \, a_n
\end{pmatrix} \, .
\end{displaymath}
The coefficient matrix of this linear system is the Vandermonde's matrix (in a non-classical form), for which is known an explicit expression of its inverse. Hence, the solutions for the unknown $\{ f_i \, a_i \}_{n \geq i > 0}$ can be worked out allowing to rewrite the approximated emission rate equation as:
\begin{equation}
   - \sum_{0 < j \leq n} \frac{1}{\lambda_j} \, \dot{\nu} 
   + \sum_{\substack{0<k\leq n \\ 0<j<k}} \frac{1}{\lambda_j \, \lambda_k} \, \ddot{\nu}
   + \cdots
   + (-1)^n \prod_{0 < j \leq n} \frac{1}{\lambda_j } \,  \partial_t^n \nu
   = \Phi - \nu
\label{eq:ApproxEREuncoupled}
\end{equation}
Note that, this ordinary differential equation (ODE) depends only on the eigenvalues $\{\lambda_n\}$ and the gain function $\Phi$. 

\begin{figure*}[htb]
\centering
\includegraphics[width=147.0 mm]{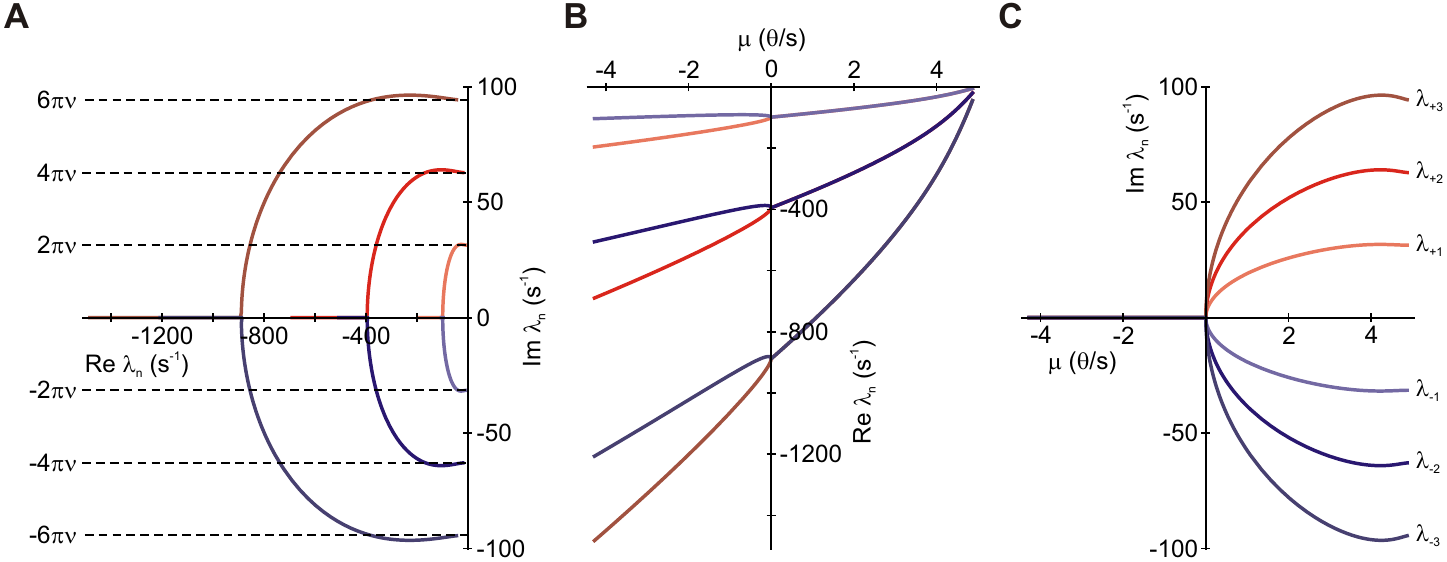}
\caption{Distribution of eigenvalues for VIF neurons. First 3 couples of modes are considered. Mean $\mu$ and standard deviation $\sigma$ of the input current are changed keeping constant the emission rate at $\nu = 5$~Hz. Darker and lighter colors are for larger and smaller $|\MyRe(\lambda_n)|$, respectively. Red (blue) curves are for $\lambda_n$ with $n>0$ ($n < 0$). {\bf A}, $\lambda_n$ distribution on the complex plane $(\MyRe(\lambda),\MyIm(\lambda))$. Eigenvalues are complex conjugates at positive drift ($\mu > 0$). Angular frequency at positive drift is expected to be $\omega_0 = 2 \pi \, n \, \nu$. {\bf B-C}, real and imaginary parts of the eigenvalues.}
\label{fig:EigenvaluesUncoupled}
\end{figure*}

Assuming an ordered spectrum of $L$, such that $|\MyRe(\lambda_n)| \geq |\MyRe(\lambda_m)|$ if $n > m$, this $n$-th order approximation of the emission rate equation is neglecting the dynamics at timescales smaller than $1/|\MyRe(\lambda_1)|^n $. As for IF neurons the eigenvalues are hierarchically distributed in couples of complex conjugates or in couples with similar real values, the second order ($n = 2$) approximation can be  effective enough:
\begin{equation}
   - \left( \frac{1}{\lambda_{+1}} + \frac{1}{\lambda_{-1}}  \right) \, \dot{\nu} 
   + \frac{1}{\lambda_{+1} \, \lambda_{-1}} \, \ddot{\nu}
   = \Phi - \nu \, .
\label{eq:2ndOrderEREuncoupled}
\end{equation}
Here, $n = +1,-1$ has been used for convenience instead of $n=1,2$. An example subset of eigenvalues is shown in Fig.~\ref{fig:EigenvaluesUncoupled} for VIF neurons, a simplified IF neuron with constant leakage and a reflecting barrier in $v = 0$ \cite{Fusi1999,Mattia2002}.

\begin{figure*}[htb]
\centering
\includegraphics[width=142.0 mm]{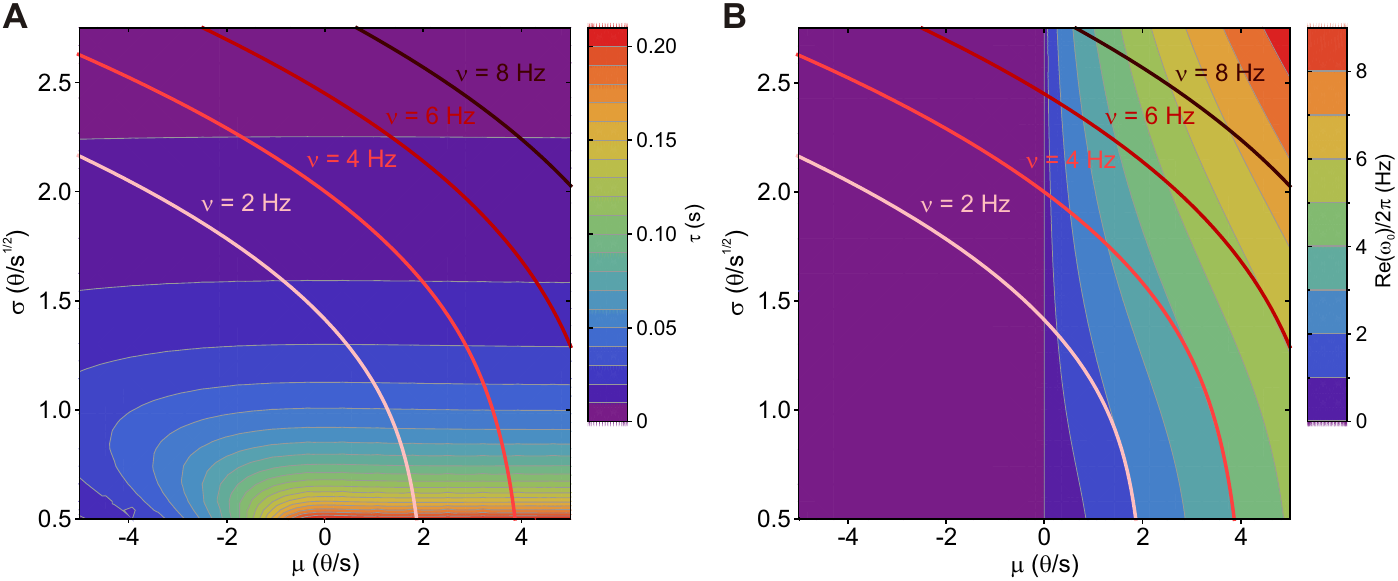}
\caption{Decay time $\tau$ ({\bf A}) and oscillation frequency $\MyRe(\omega_0)/2\pi$ ($\bf B$) in Eq.~(\ref{eq:2ndOrderEREuncoupledBis}) for VIF neurons at different $\mu$ and $\sigma$. Light and dark red lines, set of $(\mu,\sigma)$ at constant firing rate $\nu$ (see colored labels). In {\bf B}, iso-frequency curves overlap contour lines of $\omega_0$ at positive drift $\mu$ and vanishing $\sigma$, as expected.}
\label{fig:TauAndOmegaUncoupled}
\end{figure*}

For a set of uncoupled IF neurons, Eq.~(\ref{eq:2ndOrderEREuncoupled}) extends the Wilson-Cowan equation \cite{Wilson1972} in two ways: i) it is a second-order ODE such that damped oscillations can be described even when afferent currents are constant; ii) equilibrium is approached with decay time and oscillation period (if any) which depend not only on the single-neuron parameters but also on the state of the input current. This can be well appreciated rewriting Eq.~(\ref{eq:2ndOrderEREuncoupled}) as the dynamics of a damped oscillator in the force field established by $\Phi - \nu$:
\begin{equation}
\boxed{
   \tau^2 \, \ddot{\nu} + 2 \tau \, \dot{\nu}  
   = (1+\tau^2 \, \omega_0^2)(\Phi - \nu)
} \, ,
\tag{\ref{eq:2ndOrderEREuncoupled}$'$}\label{eq:2ndOrderEREuncoupledBis}
\end{equation}
where $\tau$ is the input-dependent decay time,
\begin{displaymath}
   \frac{1}{\tau} = -\frac{\lambda_{+1} + \lambda_{-1}}{2} \, ,
\end{displaymath}
and $\omega_0$ is the angular frequency of the damped oscillations (if real)
\begin{displaymath}
   \omega_0^2 = -\frac{(\lambda_{+1} - \lambda_{-1})^2}{4} \, .
\end{displaymath}
Note that this approximated dynamics is the same as the one in which only the first two non-stationary modes in the spectral expansion are taken into account, and it is equivalent to the complex-valued firing rate dynamics introduced in \cite{Schaffer2013}. An example of $\tau$ and $\omega_0$  is shown in Fig.~\ref{fig:TauAndOmegaUncoupled} for VIF neurons.

\subsection*{A network of synaptically coupled IF neurons}

A similar approach can be applied to the case of a homogeneous pool of synaptically coupled neurons. Now all functions like $\Phi$, $\vec{f}$ and $\mathbf{\Lambda}$, depend implicitly on the emission rate $\nu(t)$. This because all these functions depend on the moments of incoming currents, which in turn are modulated by the recurrent activity of the network. 

First time derivatives of the expression in Eq.~(\ref{eq:ERE}) for the stationary mode are
\begin{displaymath}
  \dot{\nu}(1-\Phi') = \vec{f}' \cdot \vec{a} \, \dot{\nu} + \vec{f} \cdot \dot{\vec{a}}
\end{displaymath}
and
\begin{displaymath}
  \ddot{\nu}(1-\Phi') -\Phi'' \, \dot{\nu}^2 = 
  \vec{f}'' \cdot \vec{a} \, \dot{\nu}^2 + 
  2 \vec{f}' \cdot \dot{\vec{a}} \, \dot{\nu} + 
  \vec{f}' \cdot \vec{a} \, \ddot{\nu} + 
  \vec{f} \cdot \ddot{\vec{a}} \, ,
\end{displaymath}
where $\Phi' = \partial_\nu \, \Phi$, $\Phi'' = \partial_\nu^2 \, \Phi$, and the same holds for $\vec{f}'$ and $\vec{f}''$.

From the expression for non-stationary modes in Eq.~(\ref{eq:ERE}), a bit lengthy formula for $\ddot{\vec{a}}$ can be worked out:
\begin{displaymath}
   \ddot{\vec{a}} = 
   \left[
      \mathbf{\Lambda}' \, \dot{\nu} + \mathbf{C}' \, \dot{\nu}^2 + \mathbf{C} \, \ddot{\nu} + (\mathbf{\Lambda} + \mathbf{C} \, \dot{\nu})^2
   \right] \, \vec{a} + 
	 (\mathbf{\Lambda} + \mathbf{C} \, \dot{\nu}) \, \vec{c} \, \dot{\nu} +
   \vec{c}' \, \dot{\nu}^2 + \vec{c} \, \ddot{\nu} \, .
\end{displaymath}

Putting together all these expressions a not so useful approximation for the emission rate equation results. Things change if a not too fast dynamics is considered. Because the adopted moving basis $\{|\phi_n\rangle\}$ rapidly adapts to the continuously varying moments of incoming currents, the hypothesis of a slow enough $\nu(t)$ dynamics allows to consider the coefficients $a_n$ of the same order of the time derivative of the input: $a_n = \mathcal{O}(\dot{\nu})$. So far, for small enough $\dot{\nu}$ a good approximation is to neglect all the nonlinear terms $\vec{a} \, \dot{\nu}$, $\vec{a} \, \dot{\nu}^2$, $\dot{\vec{a}} \, \dot{\nu}$, $\dot{\nu}^2$ and $\vec{a} \, \ddot{\nu}$. Above expressions are then simplified as follows:
\begin{displaymath}
    \ddot{\vec{a}} = \mathbf{\Lambda}^2 \, \vec{a} + 
          \mathbf{\Lambda} \, \vec{c} \, \dot{\vec{\nu}} + 
          \vec{c} \, \ddot{\nu}
\end{displaymath}
and
\begin{displaymath}
   \begin{array}{rcl}
      \dot{\nu} (1 - \Phi' - \vec{f} \cdot \vec{c}) & = & \vec{f} \cdot \mathbf{\Lambda} \, \vec{a} \\
      & & \\
      \ddot{\nu} (1 - \Phi' - \vec{f} \cdot \vec{c}) - \dot{\nu} \, \vec{f} \cdot \mathbf{\Lambda} \, \vec{c} & = & \vec{f} \cdot \mathbf{\Lambda}^2 \, \vec{a}
   \end{array} \, .
\end{displaymath}
This set of equations together with the one for the stationary mode are a linear system equivalent to the one shown in the derivation of Eq.~(\ref{eq:ApproxEREuncoupled}), and it can be solved in the same way:
\begin{equation}
\displaystyle
   - \left( \frac{1}{\lambda_{+1}} + \frac{1}{\lambda_{-1}}  \right)
   \left(1 - \Phi'- \frac{\vec{f} \cdot \mathbf{\Lambda}^{-1} \, \vec{c}}{\Tr(\mathbf{\Lambda}^{-1})} \right) \, \dot{\nu} 
   + \frac{1}{\lambda_{+1} \, \lambda_{-1}} (1 - \Phi' - \vec{f} \cdot \vec{c}) \, \ddot{\nu}
   = \Phi - \nu \, .
\label{eq:2ndOrderERE}
\end{equation}
Here a second order ODE for $\nu(t)$ is recovered as for the uncoupled case in  Eq. \eqref{eq:2ndOrderEREuncoupled}. It can be usefully rewritten as:
\begin{equation}
\boxed{
   \alpha_2(\nu) \, \ddot{\nu} + \alpha_1(\nu) \, \dot{\nu} = \Phi - \nu
} \, ,
\tag{\ref{eq:2ndOrderERE}$'$}\label{eq:2ndOrderEREBis}
\end{equation}
with
\begin{displaymath}
\begin{array}{rcl}
	\alpha_1(\nu) & = & \displaystyle
	- \left( \frac{1}{\lambda_{+1}} + \frac{1}{\lambda_{-1}}  \right)
  \left(1 - \Phi'- \frac{\vec{f} \cdot \mathbf{\Lambda}^{-1} \, 
  \vec{c}}{\Tr(\mathbf{\Lambda}^{-1})} \right) \\
	& & \\
   \alpha_2(\nu) & = & \displaystyle
   \frac{1}{\lambda_{+1} \, \lambda_{-1}} (1 - \Phi' - \vec{f} \cdot \vec{c})
\end{array}
\, ,
\end{displaymath}
where $\vec{f} \cdot \mathbf{\Lambda}^{-1} \, \vec{c} = f_{+1} \, c_{+1}/ \lambda_{+1} + f_{-1} \, c_{-1}/ \lambda_{-1}$ and $\Tr(\mathbf{\Lambda}^{-1}) = 1/\lambda_{+1} + 1/\lambda_{-1}$. Of course, in the limit of synaptically uncoupled neurons ($\Phi' = 0$ and $\vec{c} = 0$) these coefficients are the same as in Eq. \eqref{eq:2ndOrderEREuncoupled}. State-dependent decay time and angular frequency are now
\begin{displaymath}
\begin{array}{rcl}
	\tau(\nu) & = & \displaystyle \frac{2 \alpha_2(\nu)}{\alpha_1(\nu)} \\
	& & \\
   \omega_0(\nu)^2 & = & \displaystyle \frac{1}{\alpha_2(\nu)} - \frac{1}{\tau(\nu)^2}
\end{array}
\, ,
\end{displaymath}
respectively.

\begin{figure*}[htb]
\centering
\includegraphics[width=148.0 mm]{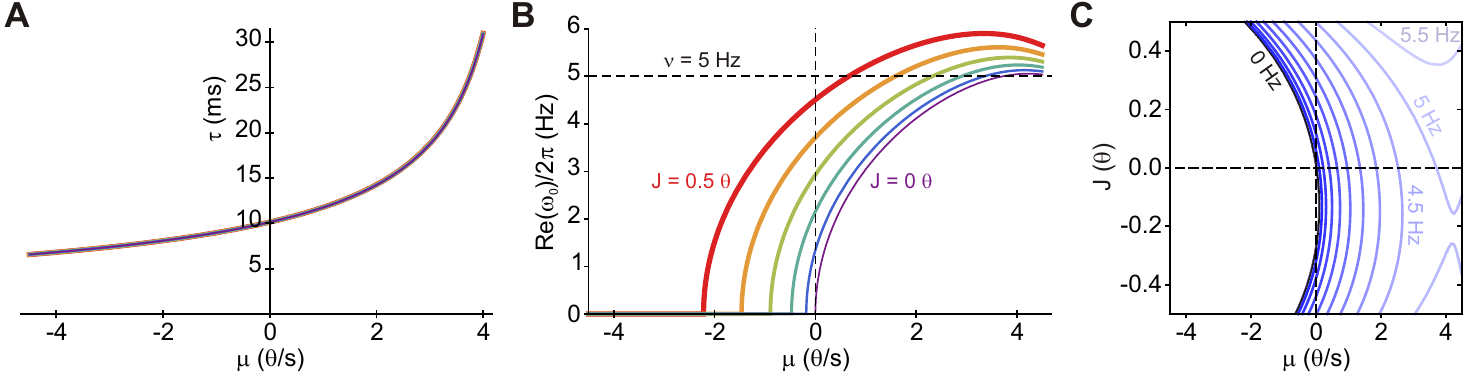}
\caption{Decay time $\tau$ ({\bf A}) and oscillation frequency $\MyRe(\omega_0)/2\pi$  ({\bf B}) versus mean afferent current $\mu$ for different synaptic efficacies $J$. As in the other figures, VIF neurons are considered. Total infinitesimal mean is assumed to be $\mu = \mu_0 + \epsilon \, J \, \nu$, where $\mu_0$ is the mean of external currents and $\epsilon = 0.2$ is the probability to have two neurons synaptically coupled. The mean-field infinitesimal variance is $\sigma^2 = \sigma_0^2 + \epsilon \, J^2 \, \nu$, where $\sigma_0^2$ is the variance of external currents. Both in panels {\bf A} and {\bf B}, $J$ is varied from $0$ to $0.5 \theta$ at steps of $0.1 \theta$ (curve colors from indigo to red, respectively). Firing rate is kept constant ($\nu = 5$~Hz) by varying $\mu_0$ and $\sigma_0^2$ accordingly. {\bf C}, contour lines of angular frequencies shown for both excitatory ($J>0$) and inhibitory ($J<0$) synaptic couplings. Damped oscillations ($\omega_0 > 0$) appear at both positive and negative drifts (black border). Contour lines correspond to constant $\MyRe(\omega_0)/2\pi$ from $0$~Hz to $5.5$~Hz at steps of $0.5$~Hz (from dark to light blue, respectively).}
\label{fig:TauAndOmegaCoupled}
\end{figure*}

In Fig.~\ref{fig:TauAndOmegaCoupled} both $\tau$ and $\omega_0$ are shown for different synaptic efficacies $J$. Infinitesimal moments $\mu$ and $\sigma$ of the input current to the VIF neurons are varied keeping constant the output firing at $\nu = 5$~Hz. Two remarks: i) $\tau$ is independent from the coupling intensity; ii) oscillatory behavior appears even at negative drift ($\mu < 0$) when $J > 0$, and it can disappear at positive drift for suited inhibitory feedbacks ($J<0$).

\begin{figure*}[htb]
\centering
\includegraphics[width=110.0 mm]{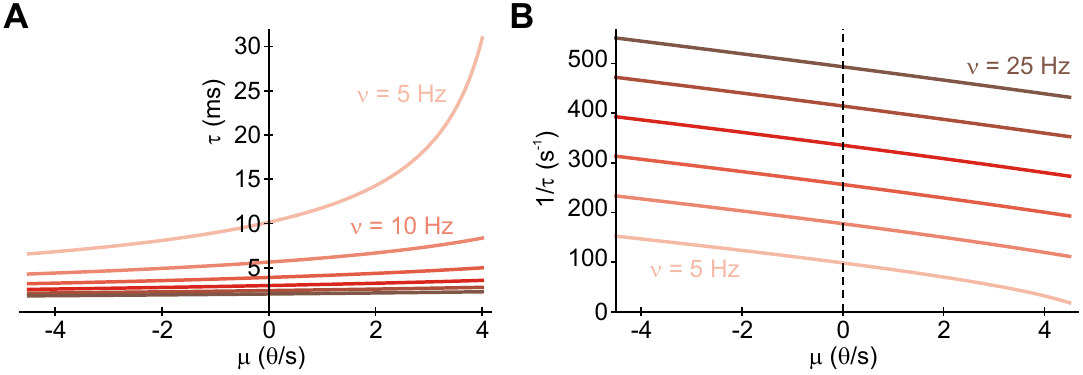}
\caption{Decay time $\tau$ ({\bf A}) versus mean current $\mu$ at different constant firing rates $\nu$. $\nu$ varies from $5$~Hz to $25$~Hz at steps of $4$~Hz (colors from light to dark red, respectively). {\bf B}, $1/\tau$ versus $\mu$, colors as in panel {\bf A}.}
\label{fig:DecayTimeVsNu}
\end{figure*}

For VIF neurons is also interesting to see how $\tau$ varies at different firing rates, as shown in Fig.~\ref{fig:DecayTimeVsNu}. Time scales are increasingly shorter for large firing rates, as expected. Furthermore, $1/\tau$ is well approximated by a linear combination of $\mu$ and $\nu$. This should not surprise because for VIF neurons $\MyRe(\lambda_{\pm1}) \simeq -2 \pi^2 \, \sigma^2 / \theta^2$, and in the plane $(\mu,\sigma)$ iso-frequency curves are rotated parabolas: $\sigma(\mu)^2 - \sigma(0)^2 \propto \mu$ (see Fig.~\ref{fig:TauAndOmegaUncoupled} and \cite{Mattia2002}). Such relationship for other kinds of IF neurons is expected to depend on the particular shape of their current-to-rate gain functions $\Phi$.

\subsection*{The case of non-stationary external currents}

The introduced approach can be used also in presence of non-stationary external currents ($\dot{\mu}_0 \neq 0$ and $\dot{\sigma}_0 \neq 0$). Considering synaptic current also due to the spiking activity of neurons outside the network, and firing at rate $\nu_{ext}(t)$, the first two moments of this additional current become $\mu_0 \to \epsilon_{ext} \, J_{ext} \, \nu_{ext} + \mu_0$ and $\sigma_0^2 \to \epsilon_{ext} \, J_{ext}^2 \, \nu_{ext} + \sigma_0^2$. What shown in the previous Section applies here provided that also the derivatives with respect to $\nu_{ext}$ are considered.

An alternative way to reduce the dimensionality of the emission rate dynamics in this case, is to rely on a more straightforward adiabatic approximation \cite{Gigante2007,Linaro2011}. For very slow external currents ($\dot{\nu}_{ext} \ll \MyRe(\lambda_{\pm 1})$), the moving basis instantaneously adapts to the time varying input, and its movement is only mildly driven by the intrinsic timescales determined by the eigenvalues of $L$. Under this hypothesis, $\dot{\vec{a}} \simeq 0$ and Eq.~\eqref{eq:ERE} reduces to
\begin{displaymath}
\left\{
\begin{array}{rcl}
   0 & = & {\mathbf \Lambda} \, \vec{a} + \vec{c} \, \dot{\nu}  + \vec{c}_{ext} \, \dot{\nu}_{ext} \\
   \nu & = & \Phi + \vec{f} \cdot \vec{a}
\end{array}
\right. ,
\end{displaymath}
where the elements of $\vec{c}_{ext}$ are obtained deriving $\langle\psi_n|$ with respect to $\nu_{ext}$ and the nonlinear terms $\dot{\nu} \, \vec{a}$ are neglected. Finally, the following first order dynamics results
\begin{equation}
\boxed{
     \tau \, \dot{\nu} = \Phi - \nu + F
} \, ,
\label{eq:1stOrderERE}
\end{equation}
where $\tau(\nu) = \vec{f} \cdot {\mathbf \Lambda}^{-1} \, \vec{c}$ is the activity dependent timescale and $F(t) = - \vec{f} \cdot {\mathbf \Lambda}^{-1} \, \vec{c}_{ext} \, \dot{\nu}_{ext}$ is an additional time-dependent forcing term. Although, all the modes of the expansion here can be considered, this approximation cannot express damped oscillations. In order words this approximation is neglecting the intrinsic diffusive dynamics related to the $\lambda_n$.

\section*{Acknowledgments}
I thank E. S. Schaffer and L. F. Abbott for a stimulating discussion from which originated my interest in the presented simplification of the FP equation. I also thank M. Augustin for a careful reading of the manuscript and the many discussions about it, and E. Hugues for an interesting feedback on the uncoupled neuron case.


\end{document}